# The Social Structure of Consensus in Scientific Review

Misha Teplitskiy, Harvard University
Daniel Acuna, Syracuse University
Aïda Elamrani-Raoult, Ecole Normale Supérieure
Konrad Körding, Northwestern University
James Evans, University of Chicago

**Corresponding authors:**

Misha Teplitskiy
Laboratory for Innovation Science
Harvard University
mteplitskiy@fas.harvard.edu

James Evans
Knowledge Lab and Department of Sociology
University of Chicago
jevans@uchicago.edu

**Acknowledgements:** We thank *PLOS* for providing data analyzed in this study. We thank participants of the Economic Sociology Working Group at MIT Sloan, Social Theory and Evidence Workshop at the University of Chicago, Society for the Social Studies of Science, Michael Menietti, Jason Radford, James Sappenfield, and Karim Lakhani for valuable insight. Any remaining errors are, of course, our own.

# Abstract

Personal connections between creators and evaluators of scientific works are ubiquitous, and the possibility of bias ever-present. Although connections have been shown to bias prospective judgments of (uncertain) future performance, it is unknown whether such biases occur in the much more concrete task of assessing the scientific validity of already completed work, and if so, why. This study presents evidence that personal connections between authors and reviewers of neuroscience manuscripts are associated with biased judgments and explores the mechanisms driving the effect. Using reviews from 7,981 neuroscience manuscripts submitted to the journal *PLOS ONE*, which instructs reviewers to evaluate manuscripts only on scientific validity, we find that reviewers favored authors close in the co-authorship network by ~0.11 points on a 1.0 – 4.0 scale for each step of proximity. *PLOS ONE*'s validity-focused review and the substantial amount of favoritism shown by distant vs. very distant reviewers, both of whom should have little to gain from nepotism, point to the central role of substantive disagreements between scientists in different "schools of thought." The results suggest that removing bias from peer review cannot be accomplished simply by recusing the closely-connected reviewers, and highlight the value of recruiting reviewers embedded in diverse professional networks.

# 1. Introduction

Around the globe, public and private organizations invest more than $2 trillion into research and development (Industrial Research Institute, 2017). Many of these organizations, including funding agencies and publishers of scientific research, face the challenging task of allocating limited financial or reputational resources across scientific projects, which require increasingly deep and varied domain expertise to evaluate (Jones, 2009; Wuchty, Jones, & Uzzi, 2007). Despite its ubiquity, however, peer review faces persistent critiques of low reliability and bias. Reviewers of a particular scientific work disagree with each other's assessment notoriously often (Bornmann, 2011; Campanario, 1998; Cicchetti, 1991; Marsh, Jayasinghe, & Bond, 2008). Indeed, agreement is often only marginally better than chance and comparable to agreement achieved for Rorschach inkblot tests (Lee, 2012). A bigger concern than the variance of reviews, however, is reviewers' bias for or against particular social and intellectual groups, and particularly those to whom they are personally connected. Given that scientists often work on highly specialized topics in small, dense clusters, the most relevant expert evaluators are typically peers of the research creators and many organizations turn to them for input. Nevertheless, many researchers have suspected that personal connections between reviewers and creators are the locus of nepotism and other systematic biases.

Several studies of scientific evaluation have demonstrated that personal connections are, indeed, associated with biased reviews. For example, recent studies document that reviewers of grant proposals and candidates for promotion favor the research of collaborators and coworkers (Bagues, Sylos-Labini, & Zinovyeva, 2016; Jang, Doh, Kang, & Han, 2016; Sandström & Hällsten, 2007; van den Besselaar, 2012). Other research documents that higher ratings tend to be given to the research of men (Bagues, Sylos-Labini, & Zinovyeva, 2017; Wennerås & Wold, 1997). These patterns of widespread disagreement and bias in scientific evaluation greatly complicate selection of the most deserving research, which generates a new problem, that of "reviewing the reviewers" to identify which of the cacophonous voices provides unbiased information or de-biasing the reviews. Meanwhile, from the perspective of scientists and scholars, evaluation decisions that drive their careers and billions of research dollars are possibly unfair and, to a large extent, the "luck of the reviewer draw" (Cole, Cole, & Simon, 1981, p. 885).

Despite the centrality of peer review to the scientific enterprise and increasing research attention devoted to it, important questions remain. First, existing studies of reviewer bias have focused on *prospective* judgments, like promotions and funding competitions. Administrators' and reviewers' task in these settings is to predict future performance. In this context, prospective judgments are inherently uncertain and may hinge on information asymmetry, such that particular reviewers have private information about the applicant that other reviewers do not possess. In contrast, it is unknown whether personal connections influence *retrospective* judgments as in manuscript review, where the task is to assess completed work. Consequently, uncertainty regarding the evaluated work should be lower and, in principle, all reviewers should have equal access to the relevant information, presented within the manuscript. In this way, it is plausible that personal connections are minimally associated with bias in retrospective evaluations.

Second, current studies do not distinguish among mechanism(s) driving bias. Reviewers may favor the work from closely connected authors for many reasons, including nepotism, similar tastes for "soft" evaluation criteria like "significance" or "novelty," and even shared views on contested substantive matters – a view we call "schools of thought" to suggest deeper shared theoretical, methodological and epistemological assumptions and commitments. Disambiguating these mechanisms is critical because the effectiveness of policies designed to mitigate reviewer bias hinge on the mechanisms driving it. In the case of nepotism, the most effective policy may be to recuse reviewers closely connected with those reviewed or provide reviewer training on conscious and non-conscious biases in judgment. In the case of soft evaluation criteria, it may be important to separate the review process into technical "objective" and softer "subjective" components. With respect to schools of thought, it may be important to select reviewers belonging to diverse schools. In practice, these mechanisms are difficult to disentangle: personal networks overlap with individuals' scientific views, and evaluations typically collapse technical and soft criteria (Lamont, 2009; Lee, 2012; Travis & Collins, 1991).

This study addresses both aforementioned shortcomings of the literature on scientific evaluation. Empirically, the study moves beyond prospective judgments and estimates the effect that personal connections play in the concrete, retrospective context of manuscript review. We use the editorial files of 7,981 neuroscience manuscripts submitted in 2011-2 to the journal *PLOS ONE*, which instructs reviewers to evaluate manuscripts' scientific validity only. We also measure connections between reviewers and authors by their relative location in the co-authorship network. Co-authorship distances are strongly related to whom authors nominate as reviewers, suggesting that formal co-authorship captures affinities between scientists. We find that reviewers give authors a ~0.11 point bonus (1.0=Reject, 4.0=Accept) for each step of proximity in the co-authorship network. Our study does not measure quality directly, so we cannot distinguish whether close reviewers *over*estimate the scientific validity of manuscripts or distant reviewers *under*estimate it. Nevertheless, if a single, true assessment of a manuscript's validity exists, our study reveals bias: reviewers' judgments systematically over- or under-shoot this value as a function of co-authorship closeness.

Conceptually, our study attempts to disentangle possible mechanisms driving co-author bias more thoroughly than prior work. To do this, we exploit the uniqueness of *PLOS ONE*'s review process and patterns in reviewer decisions. Unlike conventional journals that evaluate work on both technical and "soft" criteria, such as "significance" or "novelty," *PLOS ONE* evaluates manuscripts only on whether they are scientifically valid[1]. Reviewers disagree frequently even on this technical evaluation (inter-rater reliability = 0.19), which suggests that disagreement and biases cannot be attributed to reviewer differences on soft evaluation criteria alone. Furthermore, we find that the co-authorship bonus is *not* limited to the closest co-author connections only. Distant reviewers (co-authors of co-authors) give more favorable scores than *very* distant reviewers (co-authors of co-authors of co-authors and beyond), despite both types of reviewers having little to gain from nepotism. This pattern suggests that biases are unlikely to be driven by nepotism alone. Instead, we draw on literature from science and technology studies and sociology of science to argue that scientists' views on contested substantive matters overlap with their personal connections. Consequently, closely connected researchers are likely to belong to the same "school of thought" and favor each other's work because it matches their scientific views.

In sum, we find evidence of co-author bias in an unlikely context – judgments of scientific validity regarding completed work. The data are most consistent with scientists in the same substantive "school of thought" favoring work from other members of that school. Schools of thought problematize policies used by journals and funding agencies around the world to mitigate bias. Rather than simply recusing evaluators most closely connected from participating on the assumption of nepotism, our findings suggest that fair evaluations require evaluators from diverse scientific clusters.

## 2. Disagreement and biases in peer review

A voluminous literature has documented ways in which scientific evaluations do not necessarily converge on the underlying quality of the work or individual. Given the literature's long-standing focus on disagreement, we first establish levels of disagreement typical of conventional evaluation settings, in which validity, significance and novelty are simultaneously valued, in order to compare those levels to *PLOS ONE*, which evaluates validity alone. Next, the section reviews studies of biases in peer review associated with personal connections. We identify three mechanisms hypothesized to drive bias - nepotism, subjective review criteria, and schools of thought – and outline some conditions that mediate their salience. We return to these mechanisms in Section 4.5, which utilizes *PLOS ONE*'s uniqueness to disentangle those mechanisms more unambiguously than previously possible.

[1] *PLOS ONE* also requires that manuscripts be clearly written and adhere to the journal's data policy. A blank reviewer form is available at the following address: http://journals.plos.org/plosone/s/file?id=t6Vo/plosone-reviewer-form.pdf. Accessed 2017-12-20.

## 2.1. Empirical patterns: low reliability and favoritism

Reviewers frequently disagree about which work or person merits publication or funding (Bornmann, 2011; Campanario, 1998; Cicchetti, 1991; Wessely, 1998). Although some debate remains regarding whether peer review in multi-paradigm, low-consensus disciplines like sociology is less reliable than in high-consensus disciplines like physics (Hargens, 1988), disagreement is pervasive across disciplines (Bornmann, Mutz, & Daniel, 2010; Cole et al., 1981; Marsh et al., 2008; Rothwell & Martyn, 2000). In reviewing this literature, Cicchetti found that inter-rater reliabilities (0 = no agreement, 1.0 = perfect agreement) ranged between 0.19-0.54 for social science journals and 0.26-0.37 for medical journals (Cicchetti, 1991: Table 2). In a review of grant proposals submitted to the U.S. National Science Foundation, agreement was no better: inter-rater reliability ranged from 0.18 for chemical dynamics grants to 0.37 for economics grants (Cicchetti, 1991: Table 4). These low levels of agreement are consistent with the meta-analysis of Bornmann and colleagues, who evaluated 48 manuscript review reliability studies to find that mean inter-rater reliability had an intra-class correlation of 0.34 and a Cohen's kappa of 0.17, with the strongest studies showing lower levels of agreement (Bornmann et al., 2010). In sum, inter-rater reliabilities are low across disciplines and review settings, typically falling in the range of 0.10 – 0.40.

In addition to low reliability, a number of studies have found that reviewers favor the research of closely connected scientists. These studies typically measure personal connections by shared institutional affiliation. For example, Wennerås and Wold used data from a Swedish postdoctoral fellowship competition in medicine and found that reviewers gave applicants sharing an institutional affiliation a "friendship bonus" of 0.22 points on 0.0-4.0 scale (Wennerås & Wold, 1997). Sandstrom and Hallsten replicated this study with a newer wave of applicants to the same Swedish competition and found that applicants sharing affiliation with reviewers received a 15% bonus on scores (Sandström & Hällsten, 2007). Studies of other funding competitions have corroborated these findings. Jang and colleagues found that reviewers of grants submitted to the National Research Foundation of Korea gave slightly more favorable scores to applicants with whom they presently or previously shared an institutional affiliation (Jang et al., 2016). Bagues and colleagues (2016) found a similar pattern in promotion decisions for Italian academics: reviewers favored candidates from the same institution or with whom they had previously co-authored. Li (2017) explored the effect of a different type of relatedness on review outcomes at US National Institutes of Health funding competitions. Measuring relatedness between grant applicants and the review panel by how many members of the panel cited the applicant's work, she found that for each additional ("permanent") review member citing the applicant's work, chances of winning the grant increased by 2.2% (Li, 2017).

These studies consistently found that reviewers favored scientific work or candidates within their professional networks. In each case, however, the judgments examined were *prospective*, predicting future performance or impact. Prospective judgments are informationally distinct from retrospective ones. Prospective judgments entail uncertainty, and personal connections may give reviewers differential access to information that helps resolve uncertainty, that is whether the candidate or application will achieve its aspirations. To be sure, reviewers who personally know a candidate or a work's author(s) may draw on this private information in a biased way, as Bagues et al.'s (2016) and Li's (2017) results suggest. Nevertheless, these and other studies (Travis & Collins, 1991) indicate that information asymmetries are key to understanding the role personal connections play in prospective contexts.

An equally crucial but under-researched context is that of retrospective judgments, such as manuscript review, in which reviewers evaluate finished work. In manuscript review, information asymmetry across reviewers should be much diminished or nonexistent for two reasons. First, manuscripts should in principle report all information necessary to evaluate them. Second, in grant competitions, a small panel may be tasked with reviewing all applications from a large field, creating large disparities in how knowledgeable reviewers are about a particular application (Li, 2017). In contrast, reviewers of manuscripts tend to be selected specifically for special expertise in a manuscript's topic, resulting in relatively lower variance in reviewer expertise. Consequently, mechanisms other than information asymmetry are likely to be key in how personal connections affect retrospective judgments.

To our knowledge, only one empirical study examined personal connections in a retrospective context. Laband and Piette (1994) hypothesized that editors of economics journals use personal connections to authors in the field to "recruit" high-quality articles, by outcompeting editors without such connections. Consistent with this hypothesis, the authors found that articles published by editors who share an employing or PhD-granting institution with the articles' authors receive an average of about four more citations than articles published by unconnected editors. Beyond this example of *editor* behavior, it is unclear whether personal connections affect manuscript *reviewers'* decisions, and if so, how.

## 2.2. Mechanisms driving bias

Scholars debate the causes and epistemological legitimacy of bias in peer review (Lamont, 2009, Chapter 2; Lee, 2012). The literature has focused on three mechanisms: nepotism, subjective review criteria, and schools of thought. These are not exhaustive: reviewers may be influenced by factors that are epistemological, cognitive (Boudreau, Guinan, Lakhani, & Riedl, 2016), social psychological (Olbrecht & Bornmann, 2010; Pier et al., 2017; Roumbanis, 2016), and even ephemeral and idiosyncratic, like mood (Englich & Soder, 2009; Roumbanis, 2016). Nevertheless, each mechanism seeks to account for a major source of evaluation variation.

### Nepotism

Research on bias in peer review typically follows Robert Merton and colleagues in positing that scientific content can and should be separated from the characteristics of its creators (Robert King Merton, 1973; H. Zuckerman & Merton, 1971)[2]. When peer reviewers evaluate only qualities of the work's content, so-called

[2] This literature tends to downplay correlations between particularistic considerations and universalistic ones, yet these correlations are substantial (Lamont, 2009, p. 157). For instance, in their analysis of millions of biomedical publications, Feng Shi and colleagues find that the best predictor of collaboration (typically considered by the peer review literature as a purely "social" relationship) are similar methodological preferences (typically considered an "epistemic" preference) (Shi, Foster, & Evans, 2015).

"universalistic" considerations, peer review is deemed epistemologically legitimate and unbiased. By contrast, when reviewers consider "particularistic" characteristics of the work's creators, peer review is deemed epistemologically tainted and biased (Lamont & Mallard, 2005; Lee, Sugimoto, Zhang, & Cronin, 2013). Such considerations include anything "particular" to a works' creators and irrelevant to its content, including their gender, race, or religious identity, political ideology, any other markers of social or cultural category membership (Bagues et al., 2017), and connections between creator and reviewer.

One type of particularistic bias – nepotism – is of special concern. Scientists often work in dense clusters, and a work's most relevant reviewers are often individuals within that cluster. Collaborative connections between authors and reviewers raise the prospect that reviewers review nepotistically, favoring closely connected individuals on non-scientific grounds. Nepotism may be exacerbated by competition. Although empirical evidence within the scientific domain is anecdotal (e.g. Stumpf, 1980), studies of other domains suggest that nepotism is especially likely in competitive environments (Bazerman & Moore, 2008, pp. 156–159; Berg, 2016). The role of competition follows from a rational choice perspective: favoring personal connections on non-scientific grounds is norm-violating (Robert King Merton, 1973) and carries the risk of reputational damage and even self-deprecation. Consequently, reviewers will only take such costly action when there is something valuable at stake, such as a major competitive award. Consistent with this intuition, a survey of applicants for highly competitive grants from the U.S. National Cancer Institute found that nearly 40% viewed its peer review system as an "old boys' network," while more than 30% felt that reviewers engaged in serious norm violation of stealing applicants' ideas (Gillespie, Chubin, & Kurzon, 1985). Moreover, an experiment by Balietti and colleagues testing the effect of competition on peer evaluation of creative work found that increased competition resulted in more self-serving reviews and more frequent disagreement between reviewers (Balietti, Goldstone, & Helbing, 2016).

While evaluators closely connected to creators may have something to gain from nepotism, such as advancing the career of their students or engaging in quid-pro-quo with collaborators, the incentive for distant reviewers is much more difficult to pinpoint. Indeed, the standard policy of recusing only the most closely connected individuals, such as dissertation advisors or co-authors, from reviewing (Lamont, 2009;

Li, 2017) is presumably based on the assumption that distant reviewers have little to no such incentive for nepotism.

## Subjective criteria

A distinct collection of literature on peer review emphasizes the interpretive flexibility of "soft" evaluative criteria (Guetzkow, Lamont, & Mallard, 2004; Lamont, 2009; Lamont & Mallard, 2005). Reviewers are typically instructed to establish whether a scientific work is (1) scientifically valid and (2) "novel," "significant," or "original"[3]. There are good reasons to think judgments of significance and novelty are fundamentally more uncertain and, consequently, invite more disagreement and greater favoritism than judgments of validity. Unlike validity judgments, evaluations of impact and significance are judgments about an uncertain future: will a particular work prove valuable to a wide audience of scientists? Moreover, the concepts "significance" and "originality" are themselves ambiguous. For example, Guetzkow and colleagues (2004) found that social scientists and humanists generally agreed that significance and originality were valuable, but humanists tended to value originality regarding data or broad approach, while social scientists valued methodological and technical novelty. The editors of the high-impact biomedical journal *eLife* seek manuscripts that are "authoritative, rigorous, insightful, enlightening or just beautiful," and note that "beauty is in the eye of the beholder, and ideas about what is beautiful can change over time" (Malhotra & Marder, 2015). Furthermore, several studies find that scientists across the natural and life sciences ascribe different meanings to the terms "broader impacts" and "societal impact" (Bornmann, 2013; Derrick & Samuel, 2016; Mardis, Hoffman, & McMartin, 2012).

Multiple studies demonstrate that evaluators interpret subjective evaluation criteria in ways that favor work similar to their own (Lamont, 2009; Lamont & Mallard, 2005). For example, Travis and Collins

[3] The most prestigious scientific journals place great value on manuscripts' significance, originality, and novelty. For example, *Nature* seeks manuscripts that are of "outstanding scientific importance" and "of interest to an interdisciplinary readership," *Science* seeks manuscripts that "present novel and broadly important data," while *Cell* seeks manuscripts presenting "conceptual advances of unusual significance" on questions of "wide interest" (http://www.nature.com/nature/authors/get_published/index.html#a1, http://www.sciencemag.org/authors/science-information-authors, http://www.cell.com/cell/authors. Accessed 2017-03-13.)

(1991) relate discussions from committees evaluating physics grants, in which committee members thought it crucial to contextualize reviews from scientists with a characterization of their camps. “It’s a club . . . it doesn’t move with the times much,” wrote one committee member of the referee reports (334). When all referee reports of a particular grant originated from one “camp,” a committee member asked the organizers to obtain additional reports, as the original referees all “belonged to a mutual admiration society” (335). One evaluator from Lamont’s (2009) study of social science and humanities funding competitions lacked a close connection to an applicant, but nevertheless felt attracted to that applicant’s proposal on the basis of seemingly subjective criteria. “I see scholarly excellence and excitement in this one project on food,” the evaluator reported, “possibly because I see resonance with my own life, my own interests, who I am, and other people clearly don’t. And that’s always a bit of a problem, that excellence is in some ways … what looks most like you” (Lamont, 2009, p. 130). Another evaluator remarked, “The [proposal] on dance [I liked a lot]; I’m an avid dance person . . . in terms of studying dance, the history of dance and vernacular dance in particular. So I found that one very interesting, very good” (130).

How a bias mechanism activated by subjective review criteria affects reviewers of differing closeness to the author(s) hinges on the distribution of scientific tastes. Although quantitative data on taste distributions are lacking, examples from qualitative research including those above suggest that similar tastes are not limited to the most closely connected individuals. Neither should tastes be unrelated to closeness – proximity in social networks is consistently associated with similar tastes and opinions, whether through homophilous selection of network ties or influence between proximate nodes (McPherson, Smith-Lovin, & Cook, 2001)**.** In this way, limited current evidence suggests that similar tastes on subjective review criteria should fall smoothly with decreasing proximity or intensity of connection to author(s).

## Schools of thought

The literature summarized above tends to assume that reviewers agree on what constitutes *valid* science. They implicitly assume that a widely shared scientific method produces facts on which most can agree. The belief that a uniform scientific method compels consensus is pervasive across the physical and life sciences (Cole, 1983; Hargens, 1988) and extends to quantitative social sciences like economics (Lamont, 2009; Lazear, 2000). For example, editors of economics journals regard assessments of whether a paper is technically correct as relatively straightforward, whereas assessing its importance is “the hardest decision” (Hirshleifer, Berk, & Harvey, 2016, p. 232).

Detailed qualitative studies of scientific evaluation, however, reveal that at the research frontier, uncertainty and disagreement regarding technical matters is not unusual. Positions on such contested topics typically overlaps with scientists’ social connections (Griffith & Mullins, 1972; Travis & Collins, 1991). Communities of like-minded scholars have been analyzed at various levels of aggregation and described as invisible colleges (Crane, 1972), thought collectives (Fleck, 1979), epistemic cultures (Knorr-Cetina, 1999), paradigms (Kuhn, 1962), scientific/intellectual movements (Frickel & Gross, 2005), schools of thought (Merton 1968), and so on. We choose the term “schools of thought” for consistency with existing peer review literature (Lee, 2012) and denote by it the communities of researchers within a particular discipline that hold similar views on contested scientific topics. Shared views on contested topics rests atop shared assumptions about the world and how to properly investigate it. Recent work in the history and philosophy of science on the disunity of science highlights that topics are often contested on methodological grounds (Dupré, 1995; Galison & Stump, 1996; Geison, 1993). Methodological disagreements are ubiquitous in the humanities and softer social sciences, where different epistemic communities coexist (Abend, 2006; R. Collins, 1994; Davis, 1994; Guetzkow et al., 2004; Lamont, 2009). Nevertheless, methodological disagreements occur *at the research frontier* of even the hard sciences (Cole, 1983). Studies of peer review in the physical and life sciences reveal that when reviewers rate works separately on methodological and non-methodological aspects, agreement between reviewers on methodological aspects is no higher than agreement on significance or “reader interest” (Cicchetti, 1991; Jayasinghe, Marsh, & Bond, 2003; Lee, 2012).

Studies by scholars in the science and technology studies community further undermine the perceived straightforwardness of technical assessments (D. A. MacKenzie, 1990; Shapin, 1995). In particular, scientists do not trust all technical information equally (D. MacKenzie, 1998; Porter, 1996). For instance, in a case study on the slow acceptance of continental drift theory, Solomon (1992) demonstrated that although the quantitative evidence for the theory was widely known, how much geologists trusted that evidence depended on the geographical focus of their own work. Seemingly straightforward descriptions of scientific work, inscribed in manuscripts, are usually insufficient to overcome skepticism and diffuse complex ideas or methods (H. M. Collins, 1974, 2001; D. MacKenzie & Spinardi, 1995; Polanyi, 1958).

As with the subjective review criteria mechanism, the "schools of thought bias" varies with the proximity of connection between reviewer and author(s) and depends on the distribution of substantive scientific views. The same arguments made regarding subjective review criteria apply to schools of thought: homophily in substantive views should fall off smoothly with increasing distance between scientists.

## 2.3. Summary

Each of the three mechanisms we detail above – nepotism, subjective criteria, and schools of thought – is capable of generating bias towards close connections in peer review[4]. Consequently, the salience of each mechanism is unclear in most review settings. Distinctive peer review settings, however, can enable us to isolate mechanisms. The foregoing discussion suggests that decrease in competition should decrease bias attributable to nepotism. Furthermore, reviewers without direct connections to an applicant or author should be relatively uncorrupted by nepotism. Increase in the concreteness of review criteria, such as evaluating work on validity alone, should decrease bias attributable to subjective review criteria and increase consensus. Meanwhile, neither competition nor exclusive evaluation of scientific validity should affect the level of bias attributable to substantive disagreements between schools of thought. Prior studies have focused on prospective, highly competitive evaluations of works or individuals on the basis of validity and significance, originality, and other "soft" criteria. In such settings all mechanisms may, in principle, be equally salient. In contrast, the setting we analyze here involves retrospective evaluations of validity in an uncompetitive publishing environment. Section 4.5 utilizes the setting's uniqueness to explore the salience of each mechanism in contributing bias.

[4] The mechanisms can also generate disagreement among reviewers, if their degree of connectedness to the work's creators differs.

# 3. Data and methods

## 3.1. *PLOS ONE*

*PLOS ONE* is among the world's largest scientific journals, publishing approximately 30,000 peer-reviewed papers a year in all fields of science and medicine[5]. It was founded in 2006 by the Public Library of Science with the mission to publish and make publicly accessible all scientific work meeting high standards of scientific *validity*, regardless of the work's perceived novelty or impact[6]. "The idea … to decouple impact assessment from technical assessment," wrote Mark Patterson, director of publishing at *PLOS*, "is at the heart of the journal's novelty" (quoted in Adams, n.d.). Its first managing editor Chris Surridge stressed the problem of subjectivity:

> *Traditionally a lot of the work that goes into peer reviewing consists of asking questions like: "How significant is this? How surprising are the conclusions?" Essentially, these are subjective questions. A more objective question to ask would be: "Is this properly done science"* (interview on Poynder Blog, June 15, 2006).

Ten years later, the editor-in-chief Joerg Heber emphasized again that, "The more … journals operating without any subjective selection criteria of their published output beyond scientific validity, the better it is for science" (Heber, 2016). This claim suggests that conditional on the accuracy of scientific findings, assessments of significance and originality vary widely and should be carried out "post-publication" by the full community of scientists.

*PLOS ONE* thus welcomes types of submissions conventional publishers may reject because they are not novel or significant enough, including negative results, replications, and so on (MacCallum, 2011). Furthermore, the journal is published entirely online, enabling *PLOS ONE* to accept a nearly limitless number of submissions. Consequently, *PLOS ONE* may accept multiple articles on a given topic, easing the zero-sum competition often faced by groups that pursue similar research. The journal currently accepts approximately 70% of submissions and its 2016 impact factor[7] is 2.81. In sum, *PLOS ONE*'s validity-focused evaluation, lack of space limitations, relatively high acceptance rates, and prestige that rests well below the highest impact journals like *Science* and *Nature* reinforce each other to create a review system that prioritizes scientific validity and minimizes incentives for nepotistic reviewing.

---

[5] https://en.wikipedia.org/wiki/PLOS_ONE. Accessed 2016-10-23.
[6] http://journals.plos.org/plosone/s/criteria-for-publication. Accessed 2017-03-15.

*PLOS ONE*'s review process, like that of many other natural science journals, is single-blind: reviewers observe the identities and affiliations of the authors. Although it is intuitive to expect a single-blind review process to enable more relational bias than a double-blind process, existing studies comparing the two processes have been surprisingly inconsistent in identifying even small differences (Jefferson, Alderson, Wager, & Davidoff, 2002; Laband & Piette, 1994a; Lee et al., 2013; Rooyen, Godlee, Evans, Smith, & Black, 1998; Wessely, 1998). In at least one instance, the absence of effects cannot be explained by *unsuccessful* blinding, when reviewers are able to guess authors' identities, as successfully blinded reviewers do not appear to favor close colleagues (Justice et al., 1998). Indeed, the persistence of single-blind reviewing is based at least in part on editors' doubts that moving to a double-blind system would result in any improvements ("Nature journals offer double-blind review," 2015). It is thus unclear to what extent this feature of the review process affects our results, if at all.

Our analysis focuses on the field of neuroscience. As a life science, neuroscience does not suffer from the criticism leveled at many social science and humanities disciplines that researchers are unable to agree on fundamental assumptions and goals or to create cumulative knowledge (R. Collins, 1994; Davis, 1994; Hargens, 1988). Nevertheless, neuroscientists pursue research from multiple methods and perspectives, with familiar disagreements erupting at the boundaries between them. One axis of division concerns the appropriate level of analysis, with some neuroscientists arguing that the field's preoccupation with the characterization of individual neurons misses emergent behavior that arises when networks of neurons interact (Jonas & Kording, 2016; Krakauer, Ghazanfar, Gomez-Marin, MacIver, & Poeppel, 2017; Yong, 2017). Another axis of division concerns the degree to which findings from model organisms like mice generalize to humans (Preuss, 2000; Yong, 2017). Yet another division lies between experimentalists and computational modelers (“The practice of theoretical neuroscience,” 2005). Empirically, peer review in neuroscience displays the same (high) levels of disagreement observed in other disciplines (Rothwell & Martyn, 2000). This diversity of neuroscience raises the possibility that disagreements and biases in the review of neuroscience work reflects differences in perspectives on what type of research is scientifically valid.

---

[7] he impact factors during the period in which our data were collected were 4.09 (in 2011) and 3.73 (in 2012). Source: Clarivate Analytics. Journal Citation Reports. Accessed 2017-06-26.

## 3.2. Reviews and co-authorship

We obtained the complete review history for 7,981 neuroscience manuscripts submitted to *PLOS ONE* between 2011 and 2012. The manuscripts were authored by 46,455 individuals and reviewed by 21,665 reviewers. Many of these manuscripts went through more than one round of review, resulting in 24,022 total reviews. The sensitivity of the data necessitated strong confidentiality agreements with the journal and IRB[8]. Source data remained encrypted and noise was added to reviewer decisions—3% of decisions were randomly flipped—to enhance privacy. Only after adding noise and removing identifying information was the data analyzed.

The reviews dataset was supplemented using *Scopus*, a bibliographic database covering more than 20,000 peer-reviewed journals[9]. The full name and institutional affiliation for each author, reviewer and editor were queried with the Author Search functionality[10]. The *Scopus* API matches the name with disambiguated profiles in its database. In cases where a search returned multiple results with unrelated information, we performed a secondary search on these results to compute simple edit-distance matching. After obtaining the *Scopus* ID of all authors, we used another call to the API to obtain their *h*-index, or the maximum number of articles *h* that each scientist had published receiving *h* or higher citations. While

[8] IRB was granted through one of the authors’ institutions and only two authors had access to source data.
[9] https://www.elsevier.com/solutions/scopus/content. Accessed 2017-07-19.
[10] https://dev.elsevier.com/sc_apis.html. Accessed 2017-03-15.

*Scopus* does not capture all scientific output for authors, largely lacking conference proceedings, proceedings are not viewed as a critical output in neuroscience departments, and the *h*-index was only used for relative comparison between authors.

A *Scopus* query[11] was used to identify the lifetime co-authors of each reviewer, author, and editor. Resulting co-authorship ties were combined into a single network. The final co-authorship network included 1,822,998 individuals (PLOS authors, reviewers, editors, and their co-authors) and 4,188,523 co-authorship ties. In this network, the average neuroscientist has 4.59 co-authorship connections, 3,258 neuroscientists have 0 co-authors, and 1,124,153 neuroscientists have 1 co-author. Individuals with only 1 co-authorship tie tend to be co-authors of the *PLOS* individuals whom we initially queried[12].

This co-authorship network comprised of *PLOS* individuals and their "one step" connections is remarkably well connected: 99.6% of the ~1.8M individuals are connected directly or indirectly, with very few singletons or separate clusters. Nevertheless, the network is measured only partially, as *Scopus* query limits made it infeasible for us to identify all second order (co-authors of co-authors) and higher order connections. Consequently, measured distances between any two individuals must be interpreted carefully. Details on measurement error are located in Supplementary Materials A. Ties of length 1, 2, and 3 should not be affected by artificial sparseness. Ties of length 4 or higher, however, are likely artificially high—some would be of length 3 in the complete network. To reflect accurate uncertainty about the distances of ties measured as length 3 or greater, we relabel all such ties as "3+."

Manuscripts were typically authored by several individuals. In order to measure a reviewer's co-authorship distance to a *manuscript*, it was necessary to aggregate the several reviewer-author distances into a single number. Figure 3.1. illustrates this issue for both co-authorship and expertise.

---

[11] This query was capable of returning at most 179 co-authors with whom the individual collaborated over her lifetime. The degree and consequences of this censoring are discussed in Supplementary Materials A.

[12] This network is thus a "one step out" network - it begins with "seed nodes" from *PLOS ONE* and measures their connectivity to all other nodes, but does not measure the connectivity of these other nodes, as a "two steps out" network would. Consequently, the measured network is sparser than in reality.

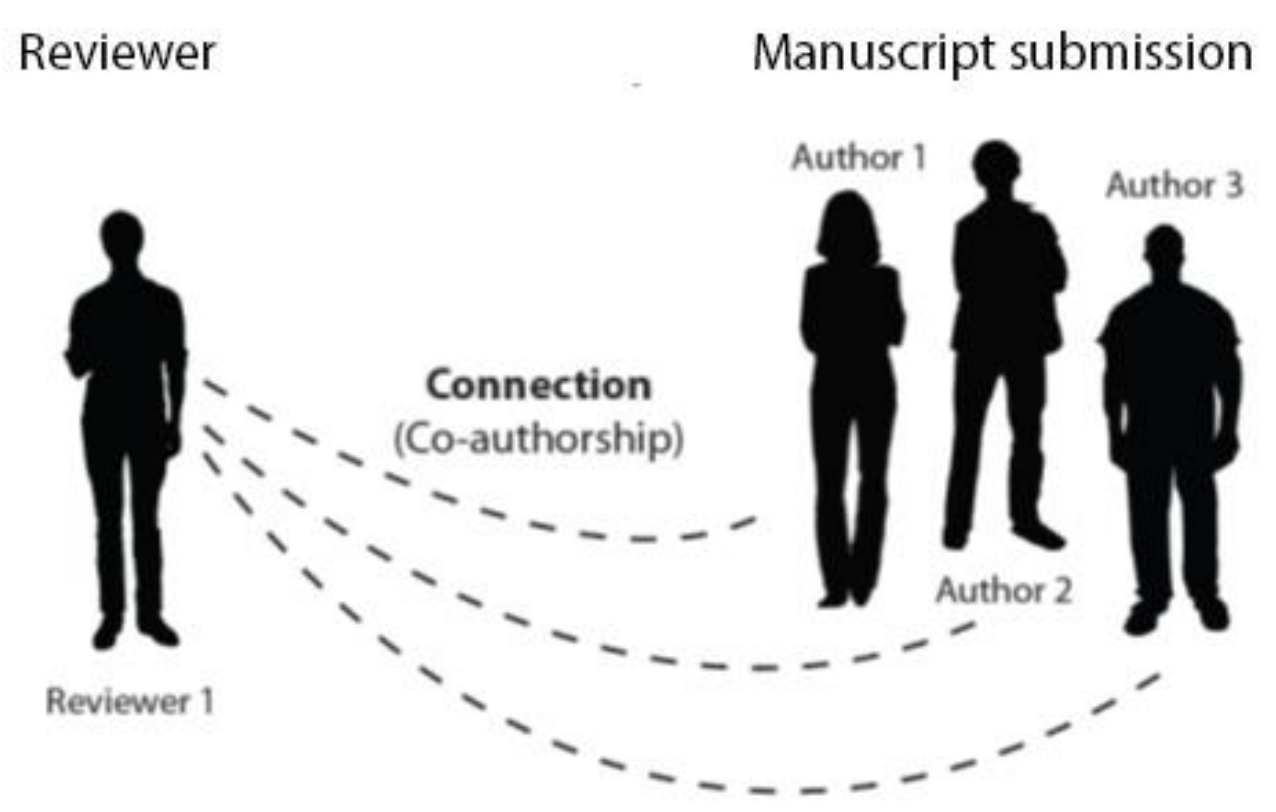

**Figure 3.1.** Diagram illustrating the measurement of co-authorship connection between a reviewer and author(s) of a manuscript. To aggregate the multiple co-authorship measures we take the minimum distance (e.g. closest co-authorship between a reviewer and any of a manuscript's authors).

The hypothetical manuscript submission has three authors, so Reviewer 1 has three separate and co-authorship distances to the manuscript. Intuitively, we could aggregate these distances with minimum, maximum, mean, or mode. In our context, minimum distance makes the most sense because lab-based research is common in neuroscience, and so a reviewer may have a close tie to the principal investigator heading the lab, but a distant tie to the manuscript's other, more junior authors. To the extent that principal investigators personify the cognitive and social dimensions of the lab's output, minimum distance between a reviewer and all authors would most accurately measure the connection between reviewer and manuscript. Figure 3.2. displays the distribution of minimum co-authorship distance.

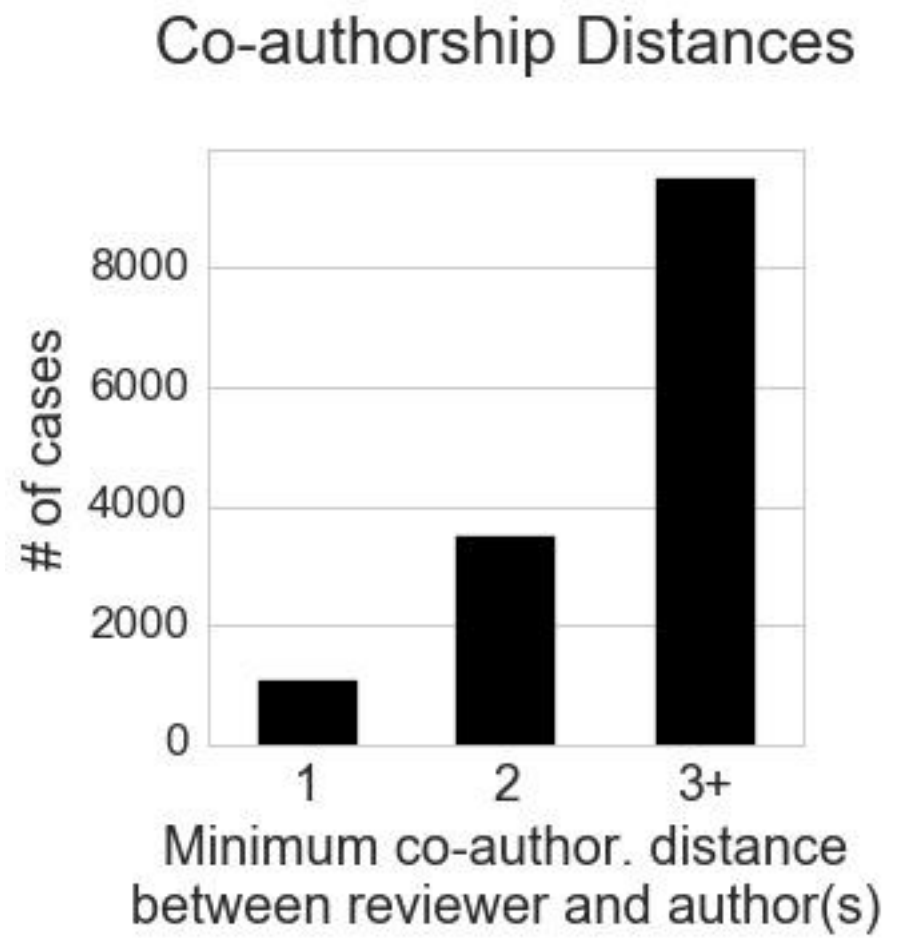

**Figure 3.2.** Histogram of minimum co-authorship distances between manuscript reviewers and authors. Distance 1 denotes co-authors, 2 denotes co-authors of co-authors, and so on.

Close co-authorship ties (distance 1 and 2) are infrequent relative to 3+ ties. Nevertheless, it is notable that more than 1000 reviews in the dataset were written by reviewers who had at one time co-authored with at least one of the authors.

To assess whether formal co-authorship connections are meaningful signals of personal or intellectual affinity, we compared them with choices *PLOS* authors faced when submitting manuscripts: who to nominate as a reviewer[13]? Although we are not aware of systematic evidence on how authors choose suggested reviewers, it is plausible that they nominate individuals most likely to provide a favorable review, perhaps with the additional constraint that the nomination does not violate obvious conflicts of interest. The set of nominated reviewers may be considered an "ideal" personal network – individuals *most likely to like* each other's work. To the extent that this privately nominated "ideal" network overlaps with the actual network constructed from co-authorship connections, the co-authorship network serves as a meaningful proxy of affinities between scientists. Figure 3.3 displays the overlap between real and ideal networks.

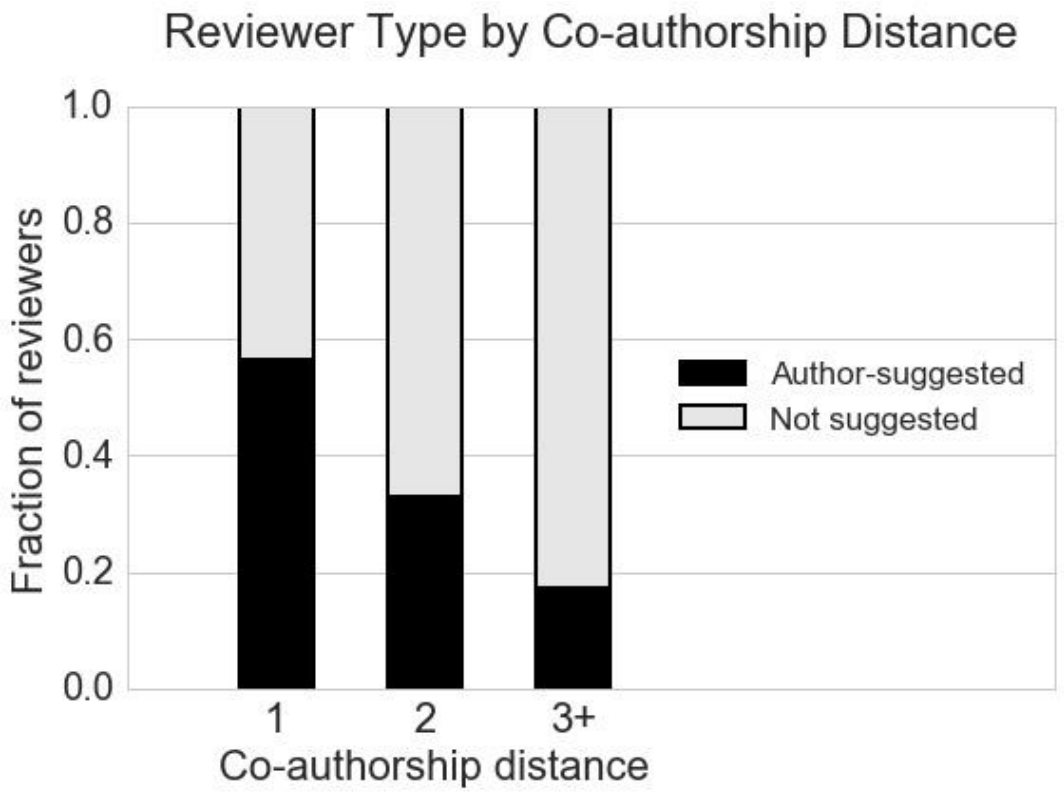

**Figure 3.3.** Distribution of author-suggested and non-suggested reviewers by their distance to the author(s). Co-authorship distance and nomination are meaningfully associated (ρ = -0.27) with closer reviewers likelier to be nominated.

The figure shows a clear pattern: the more distant the reviewer, the less likely he or she to have been author-nominated. Meanwhile, more than 50 percent of the closest reviewers (co-authors) were nominated by the author(s). Although analyses comparing reviews from nominated to non-nominated reviewers have been published, to our knowledge this is the analysis of *who* is nominated in the first place, and the first to establish the usefulness of easily observable co-authorships as a proxy the typically unobserved affinities between scientists' views.

[13] *PLOS ONE* discontinued the practice of allowing reviewer recommendations in 2014.

## 3.3. Measuring disagreement

We measured disagreement between reviewers separately in each round of review for two reasons. First, inter-rater reliability metrics assume that raters make decisions independently (Hayes & Krippendorff, 2007), but *PLOS ONE* reviewers observe each other's decisions after the first round, so decisions in subsequent rounds are likely influenced by this information. Second, it is unclear whether the goals of later rounds of review match those of the first. Evidence is limited, but in conventional journals, review of revise-and-resubmit appears to focus narrowly on the altered scope or rhetorical framing (Goodman, Berlin, Fletcher, & Fletcher, 1994; Strang & Siler, 2015a; Teplitskiy, 2015).

We used Krippendorf's alpha as the index of inter-rater reliability[14] (Hayes & Krippendorff, 2007). Although the literature on inter-rater agreement uses a wide variety of measures (Hallgren, 2012; Krippendorff, 2004), Krippendorf's alpha has three advantages: it allows for ordinal ratings, a variable number of reviewers per manuscript, and it measures agreement above the level expected from random[15] decisions. An alpha of 1.00 denotes perfect agreement, while 0.00 denotes no statistical relation between manuscript and ratings. The worst-case scenario of 0.00 can be interpreted as one in which "coders do not understand what they are asked to interpret, categorize by throwing dice, or examine unequal units of analysis, causing research results that are indistinguishable from chance events" (Krippendorff, 2004, p. 413). The rating scale is treated as ordinal, which penalizes disagreements on adjacent rating categories less than distant ones. For instance, two reviewers who assess a manuscript as “Accept” and “Minor revisions” are assumed to have less disagreement than if they evaluated it as “Accept” or “Reject.”

---

[14] Additionally, we report intra-class correlations and Cohen's weighted Kappas for comparability with other studies.

[15] In the context of inter-rater reliability, random decisions are conceptualized as follows: reviewers are assumed to know the overall probability distribution of decisions (e.g., 90% Rejects, 10% Accepts) and select their decision randomly from that distribution. Indices of inter-rater reliability measure agreement levels above the baseline level expected from such random decisions.

## 3.4. Modeled variables

**Dependent variable: *review score*.** Reviewers were instructed to provide a written report and assign the manuscript one of four recommendations: “Accept,” “Minor revision,” “Major revision,” and “Reject.” For ease of analysis, these recommendations were recoded to numerical scores of 4.0, 3.0, 2.0, and 1.0, respectively.

**Explanatory variables.** The central explanatory variable is co-authorship distance between reviewer and author(s). Two additional variables are included to account for the prominence or seniority of reviewers and authors. Controlling for prominence is important because it is correlated with network distance (see Supplementary Materials B), but prominent reviewers may review in ways not accounted for by the mechanisms on which this study focuses. For example, prominent reviewers may delegate reviews to their students or postdocs or they may take on the perspective of a “spokesperson” for a particular sub-field. To account for a prominence effect, we include the reviewer’s *h*-index and her total number of connections in the co-authorship network. For regression specifications without manuscript fixed effects, we include the author(s)’s mean *h*-index and number of network connections. Table 3.2 describes these variables; Supporting Materials contains the complete correlation matrix.

**Table 3.2.** Descriptive statistics for variables used in analysis.

| | Description | Mean | SD | Min. | Max. | Valid obs. |
|---|---|---|---|---|---|---|
| *Review score:* | | | | | | |
| *all rounds* | Reviewer's decision, coded as 1.0= Reject, 2.0=Major revision, 3.0=Minor revision, 4.0=Accept | 2.59 | 1.03 | 1.0 | 4.0 | 24022 |
| *1st round only* | | 2.24 | 0.86 | 1.0 | 4.0 | 16085 |
| *Co-authorship distance* | Minimum distance in co-authorship network between reviewer and manuscript author(s) | 2.60 | 0.63 | 1 | 3 | 14090 |
| *Num. of ties of reviewer* | Degree of reviewer in co-authorship network | 98.96 | 74.46 | 0 | 401 | 16600 |
| *Mean num. of ties of author(s)* | Mean degree of manuscript author(s) in co-auth. network | 84.02 | 48.52 | 0 | 412 | 23163 |
| *h-index of reviewer* | Hirsch index of reviewer | 18.29 | 14.70 | 0 | 132 | 16263 |
| *Mean h-index of author(s)* | Mean Hirsch index of manuscript author(s) | 14.76 | 9.83 | 0 | 101 | 23104 |

*Note:* Because of measurement limitations, co-authorship distances >3 are recoded as 3.0 and referred to as "3+" (see Supplementary Materials A).

# 4. Results

This section is divided into five parts. The first describes the overall distribution of review decisions and the second focuses on disagreement between manuscript reviewers. The third details the effect of co-authorship connections on review decisions and the fourth estimates regression models that control for the manuscript characteristics. The final part focuses on disambiguating mechanisms driving co-author bias.

## 4.1. Decisions across review rounds

Of the 7,981 neuroscience manuscripts submitted for an initial round of review, 73% were eventually accepted. 51% of the manuscripts received a final decision during the initial round of review, 40% continued on to a second round, and 8% went for a third round. Figure 4.1 displays the distribution of the 24,022 total review decisions across the first four rounds of review[16].

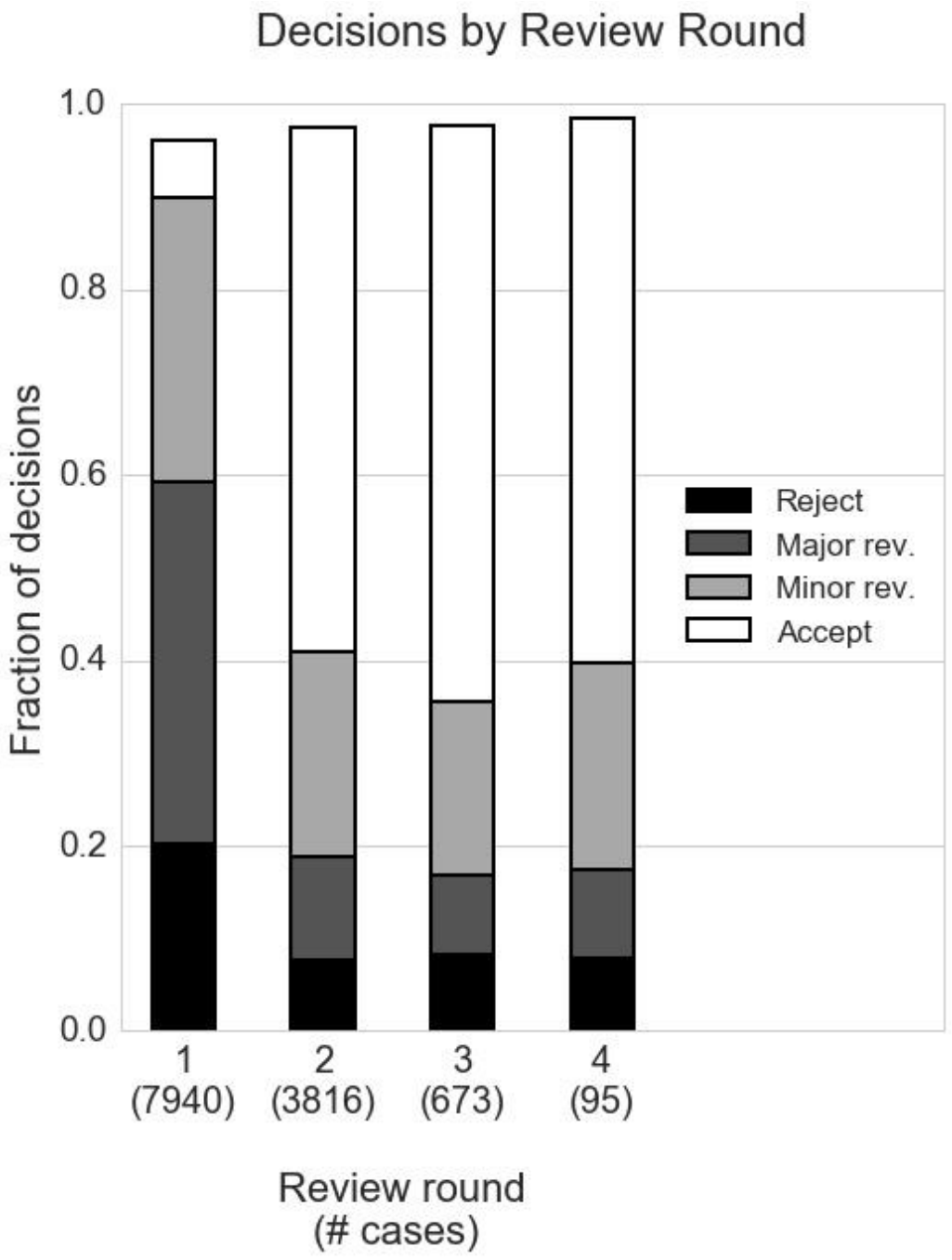

**Figure 4.1.** Distribution of decisions across review rounds (num. of cases in parentheses). Manuscripts undergoing more than one round of review are included in the calculations for each relevant round. Fractions in each round do not sum to 100% because a small fraction of submissions is terminated by the author(s) or the editor(s). A negligible fraction of manuscripts underwent more than 4 rounds of review.

Figure 4.1 shows that reviewers make use of the entire range of decisions, particularly in the first two rounds. “Accept” decisions are relatively rare in the first round, but become more frequent in subsequent rounds; “Rejects” follow the opposite pattern. The distribution of decisions across rounds is consistent with revise-and-resubmit trajectories observed in other journals. In particular, studies of journals in sociology, management, and medicine find that (1) reviewers challenge manuscripts in the first round and (2) authors attempt to address the challenges, particularly to revising the manuscript’s interpretation and scope (Goodman et al., 1994; Strang & Siler, 2015b; Teplitskiy, 2015). If reviewers in the second and subsequent rounds perceive that challenges are sufficiently addressed, the manuscript is accepted. In the analyses that follow, we analyze decisions across rounds separately for two reasons. As described above, studies of revise-and-resubmit practices indicate that the function of review may change qualitatively across rounds, from putting forth challenges in the first round, to assessing compliance in second and subsequent rounds. Reviewers typically observe each other's evaluations after the first round, weakening the assumption of independence across reviewers.

[16] A small number (<1%) of manuscripts went through more than 4 rounds of review.

## 4.2. Within-paper disagreement

Here we describe within-paper disagreement – inter-rater reliability (IRR) – of reviewers, across rounds. We measure IRR with Krippendorf's alpha (see Section 3.3), which equals 0.0 when there is no statistical association between the decisions of distinct reviewers for a given manuscript and 1.0 with perfect agreement[17]. Figure 4.2 displays inter-reviewer reliability observed in the first four rounds of review.

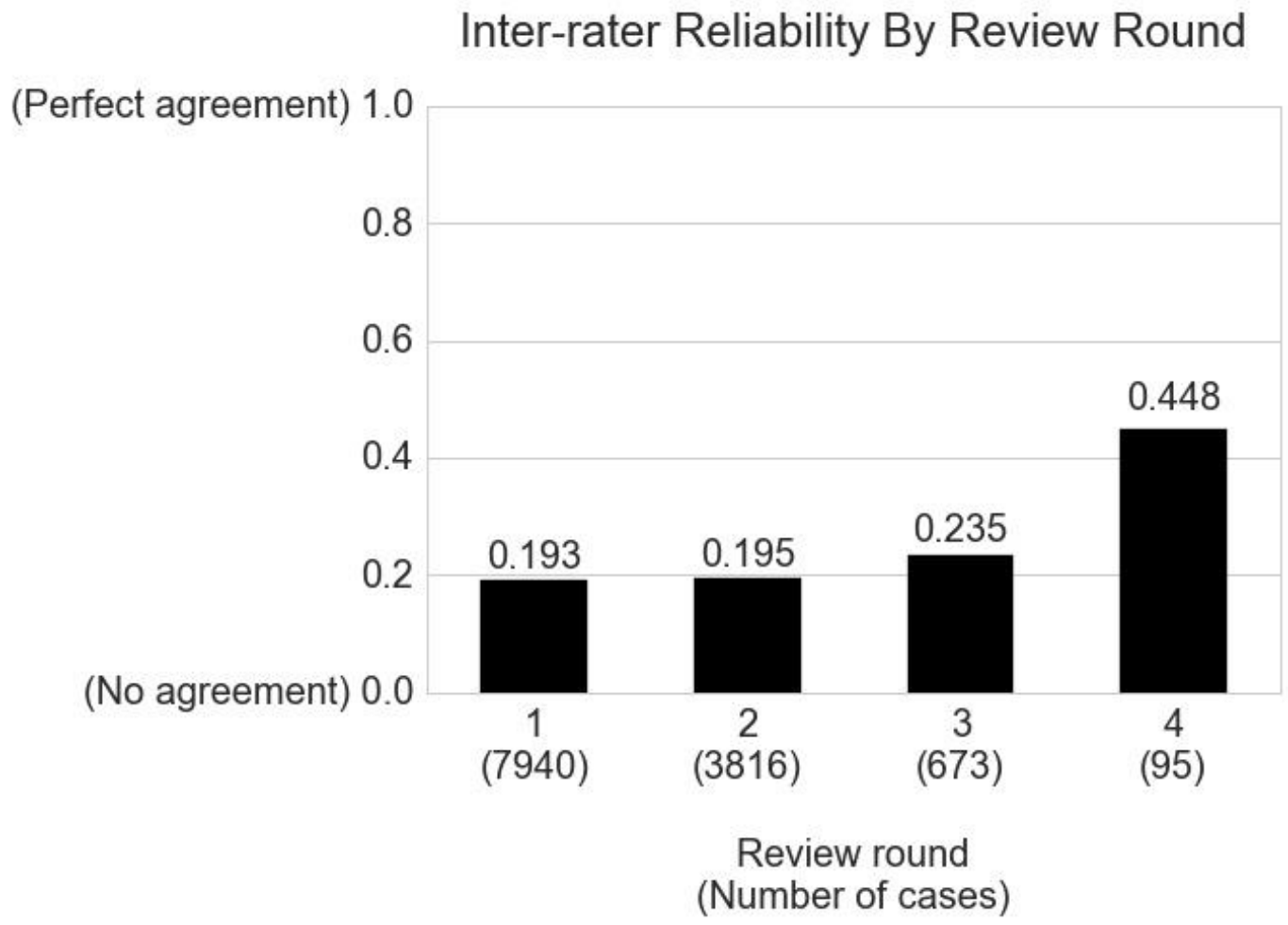

**Figure 4.2.** Inter-rater reliabilities (IRR) by review round. IRRs were computed with Krippendorf's alpha and assume an ordinal level of measurement. Note: computation of IRR in second and later rounds does not meet the assumption of independence between raters, as reviewers observe and may be influenced by each other's response following the first round.

[17] To improve comparability with existing studies (e.g. Cicchetti, 1991), we also report the intra-class correlation coefficient ICC (computed after nominal decisions had been converted to numeric values between 1.0 = Reject and 4.0 = Accept) and Cohen's weighted Kappa. Because both metrics require an equal number of reviewers for each paper, which in our reviews dataset is not always the case, we only include the first 2 reviewers in the computations. ICCs in the first four rounds of review were 0.192, 0.194, 0.204, and 0.555. Weighted Kappas were computed with squared weighting that penalizes distant mistmatches between reviews ("Accept" vs. "Reject") more than proximate mismatches ("Accept" vs. "Minor revision"). Kappas in the first four rounds were 0.192, 0.194, 0.206, and 0.545.

Inter-reviewer reliability is poor throughout the review process, but particularly in the initial rounds of review. Indeed, reviewers' decisions in the first round are perilously close to bearing no relationship with one another – IRR is 0.193. Interestingly, inter-rater reliability does not increase appreciably in round two or even three. This pattern indicates that some fraction of manuscripts in the review pipeline are controversial and these contentions are not easily resolved through revision between rounds. Manuscripts on which reviewers agree exit the pipeline toward rejection or publication along the way.

Overall, levels of disagreement for validity-oriented review are entirely comparable to those observed in more conventional review settings that value subjective impressions of "impact" and "novelty". Indeed, our IRR of 0.193 is strikingly similar to the average value of 0.17 that Bornmann and colleagues (2010) obtained in a meta-analysis of 48 studies of manuscript peer review. Insofar as reviewers follow the *PLOS ONE* mandate to evaluate accuracy alone, low levels of observed agreement here problematize the perspective that disagreement in conventional review decisions are explained primarily, if at all, by subjective review criteria. Instead, the observed pattern is consistent with the perspective that substantive disagreements between schools of thought are widespread and likely account for much of what has previously been attributed to nepotism and subjective review criteria.

## 4.3. Co-authorship connections and review outcomes

Here we describe the relationship between reviewers' decisions and co-authorship connections between authors and reviewers. Figure 4.3 illustrates the distribution of reviewers' decisions as a function of their co-authorship distance to manuscripts' authors. The left panel displays this distribution for the first round of review and the right displays it for the second and subsequent rounds.

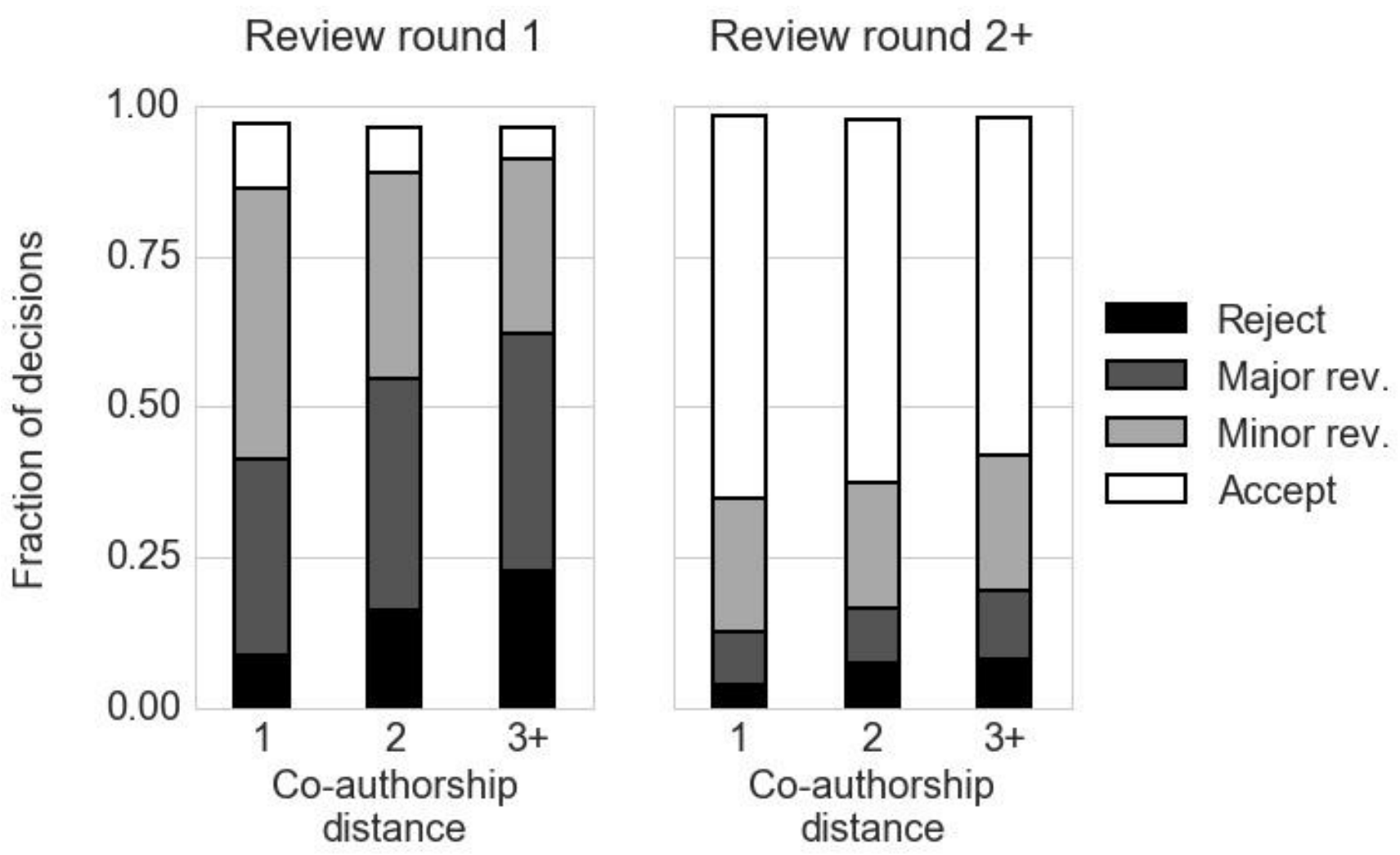

**Figure 4.3. Left.** 1st round review decisions by reviewer's (minimum) co-authorship distance to manuscript's author(s). **Right.** 2nd or later round review decisions by reviewer's (minimum) co-authorship distance to manuscript's author(s). In both panels, the farther the author is from the reviewer, the harsher the decision. Fractions at each distance do not sum to 100% because a small fraction of submissions is terminated by the author(s) or the editor(s).

Both panels show that each additional step of distance in the co-authorship network is associated with harsher reviews: "Rejects" and "Major Revisions" increase, while "Accepts" and "Minor Revisions" decline in frequency. Because we observe these patterns across manuscripts, however, it is possible that reviewer distance is correlated with manuscript characteristics, e.g., editors assign manuscripts of poorer quality to more junior, distant, or otherwise peripheral reviewers. Conversely, the literature on categorization and valuation provides a plausible alternative mechanism. This literature found that when creative and technical work falls squarely within well-established paradigms, quality judgments quality become cognitively easier (Hsu, Roberts, & Swaminathan, 2011; E. W. Zuckerman, 1999) and relatively error free (Ferguson & Carnabuci, 2017; Leahey, Beckman, & Stanko, 2017). Scientific work that spans boundaries may be more difficult for editors to "place," resulting in assignment to ill-fitting and possibly harsher reviewers.

To explore the possibility that reviewer co-authorship distance is correlated with unobserved manuscript characteristics, we compare decisions from reviewers who assess the same manuscript. Figure 4.4 displays differences in within-manuscript decisions during the first round of review (left panel) and second and later rounds (right panel).

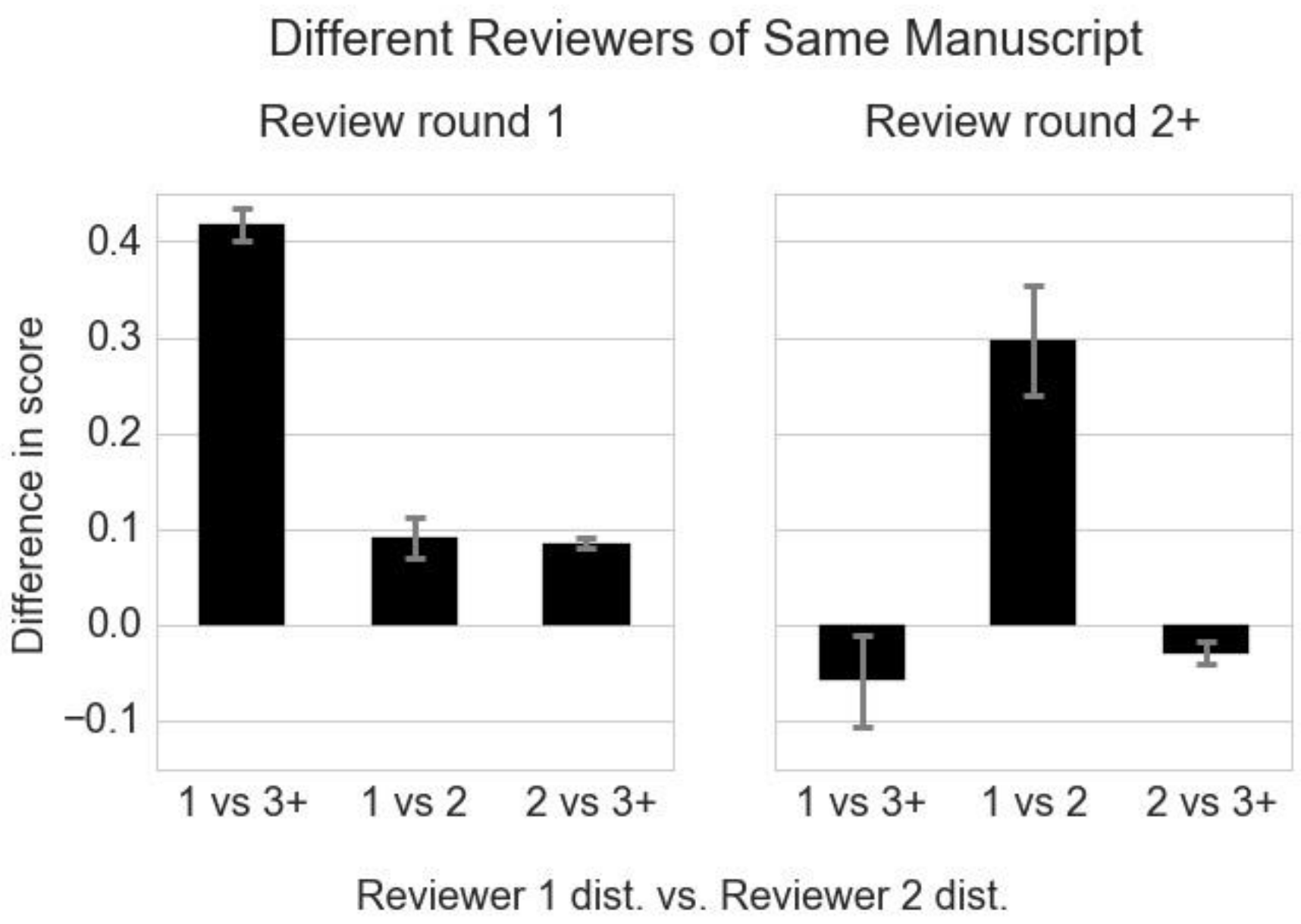

**Figure 4.4. Left.** Differences in 1st round decisions between reviewers who are close or far from the manuscript's author(s). Reviewers favor authors closer in co-authorship. As expected, the biggest difference (0.42 points) is observed in maximally close (1) and distant (3+) review panels. **Right.** Differences in decisions during 2nd and later review rounds. In these rounds, the only substantial difference is between the increased favor shown by reviewers of distance 1 rather than 2. In both panels, decisions were converted to numerical scores as Reject=1.0, Accept=4.0.

Within-manuscript patterns that we observe are consistent with those found across manuscripts: reviewers favor closer co-authors. In the initial round of review, reviewers who had at some point co-authored with one or more authors of the manuscript (distance 1) scored manuscripts an average of 0.42 points better than reviewers who are of co-authorship distance 3 or higher. This difference represents approximately 40% of a decision "level " *e.g.* between *Major revision* and *Reject*, and about 49% of the standard deviation of first-round review scores (*sd* = 0.86). This level of favoritism is of the order observed in settings where science or scientists are encouraged to evaluate on "importance" criteria beyond validity, *e.g.* 0.22 points on a 5.0-point scale (Wennerås & Wold, 1997).

Differences in decisions shrink with the difference in co-authorship distance between author(s) and reviewers. Reviewers who are co-authors of co-authors (distance 2) offer scores on average 0.09 more favorable than reviewers of co-authorship distance 3+. The pattern observed in rounds 2 and later is less unequivocal. Here, the main difference is between close (distance 2) and very close (distance 1) reviewers:

more proximate reviewers give a bonus of 0.30. Surprisingly, in review panels comprising other types of reviewers, more distant reviewers give more favorable scores than close ones, albeit by a tiny amount (< 0.06 points). In sum, decisions across and within papers, particularly in the initial round of review, reveal that reviewers favor close authors in the co-authorship network.

## 4.4. Regression models

Here we focus on isolating the effect of co-authorship from manuscript and reviewer characteristics. Co-authorship distances are negatively correlated with reviewers' prominence as measured by the $h$-index ($\rho = -0.257$, $p < 0.001$): more prominent reviewers are more closely connected to the authors they review. It is thus desirable to control for reviewers' $h$-indices and overall network connectivity to understand whether apparent effects from co-authorship connection hold independent of these quantities.

We focus on review decisions in the first round to best meet the assumption of independence between observations and because subsequent rounds likely differ in function from the first. We estimate a set of linear models with the following specification:

$$REVIEW.SCORE_{ij} = FE_j + \beta_1 COAUTHOR.DIST_{ij} + \beta_2 CONTROLS_{ij} + REVIEW.ROUND_{ij} + \epsilon_{ij}$$

Fixed effects $FE_j$ for each manuscript in the first round of review control for unobserved manuscript and author characteristics, such as inherent quality, and enable us to isolate effects of co-authorship more directly than studies that use group, rather than individual, review decisions (*e.g.* Wennerås & Wold, 1997). The dependent variable $REVIEW.SCORE_{ij}$ is the numerical review decision (1.0=Reject, 4.0=Accept) by reviewer *i* of manuscript *j*. $COAUTHOR.DIST_{ij}$ is the minimum co-authorship distance between reviewer *i* and manuscript *j*'s author(s) and takes on the values 1, 2, or 3. The $CONTROLS_{ij}$ vector includes reviewer *i*'s $h$-index and her overall number of co-authorship network connections. Although the focus is on first-round decisions, $REVIEW.ROUND_{ij}$ is added for the last model, which pools data from all rounds. Table 4.1 presents estimates from partial and full regressions specified above.

**Table 4.1.** Regressions estimating the associations between reviewers' score deviations with co-authorship, expertise, and controls.

| | Dependent variable: *Review score* (1.0=Reject, 4.0=Accept) | | | | |
|---|---|---|---|---|---|
| | (1) | (2) | (3) | (4) | (5) |
| Co-author distance (1=close, 3+=far) | -0.195*** (0.014) | -0.219*** (0.015) | -0.088*** (0.025) | -0.107*** (0.026) | -0.101*** (0.021) |
| *Round of review* | | | | | 0.864*** (0.014) |
| Controls: | | | | | |
| *Reviewer* h-*index* | | -0.0003 (0.001) | | -0.001 (0.002) | 0.0003 (0.001) |
| *Reviewer* n. *of co-authorships* | | -0.001*** (0.0002) | | -0.001* (0.0004) | -0.001*** (0.0003) |
| *Author(s)'s mean* h-*index* | | 0.007*** (0.001) | | | |
| *Author(s)'s mean* n. *of co-authorships* | | -0.001*** (0.0003) | | | |
| Manuscript FE | N | N | Y | Y | Y |
| Observations | 9,092 | 9,032 | 9,092 | 9,040 | 13,580 |
| $R^2$ | 0.021 | 0.027 | 0.004 | 0.007 | 0.356 |
| F Statistic | 190.806*** (df = 1; 9090) | 49.846*** (df=5; 9026) | 11.916*** (df = 1; 2998) | 4.337*** (df = 6; 2964) | 1027385*** (df = 4, 7447) |

*Note:* Estimates from OLS regressions. The dependent variable is Review score, which takes on values 1.0=Reject to 4.0=Accept. Asterisks *, **, and *** indicate significance at the $p < .1$, $p < .05$, and $p < .01$ levels, respectively, with 2-tailed t-test. The $R^2$ for models with fixed effects is the within-manuscript variance explained.

Baseline Model 1 regresses first round review score on co-authorship distance. Estimates from model 1 reveal that close co-authorship connections are associated with favorable reviews: a one-step increase in co-authorship closeness between the reviewer and the manuscript author(s) is associated with a 0.195-point improvement in review score. This effect represents 23% of a standard deviation for first-round review scores. In Model 2, which adds several reviewer and author(s) controls, co-authorship distance remains significant and increases in size.

Model 3 regresses review score on co-authorship with manuscript fixed effects. The effect of co-authorship distance is in the same direction but nearly halved in size: a step of proximity is now associated with only a 0.088-point improvement in review score – 10% of a standard deviation for first-round scores. The decrease in effect size suggests that higher quality manuscripts, as measured by favorable review scores, tend to be assigned to closer reviewers. Alternately, editors may have an easier time "locating" manuscripts that fit neatly into conventional research paradigms, and assign them to reviewers familiar with and sympathetic to those paradigms. The change in effect size between random and fixed effects models underscores the importance of fully controlling for observed and unobserved manuscript characteristics. Model 4 adds reviewer controls to the fixed effects model, leading to a slightly increased co-authorship effect size (12% of a standard deviation for first-round scores). Model 5 pools data from all rounds of review and is presented for completeness. As expected, reviewers' scores are substantially higher in later rounds of review, and the effect of co-authorship does not appreciably change.

The variance in review scores explained by our four main models, particularly fixed effects models 3 and 4 (see Table 4.1, note), is modest. It is instructive to compare these estimates to a comparable study of scientific evaluation. Bagues and colleagues' (2015) studied promotion decisions in Italian academia. The authors found that connections between evaluators and candidates explain 0.2% of the variance in evaluators' decisions, accounting for candidate fixed effects (Bagues et al. 2015, Table 7, model 1). The explained (adjusted) variance increases to 6.6% when they add evaluator fixed-effects (Bagues et al. 2015, Table 7, model 2), which our analysis does not do. Considered together, these studies highlight just how poorly variation in evaluation decisions is understood, even though specific evaluators score consistently across papers. Accounting for connections, one of the most intuitive sources of variation, leaves much to explain.

Lastly, "Reject" is the most consequential recommendation a reviewer can make. In order to explore how rejection probability varies with reviewer distance, we recoded reviewers' decisions as 1.0=Reject and 0.0=otherwise. Results from a logistic model[18] of rejection on co-authorship distance, reviewer's *h*-index, number of co-authorship ties, and manuscript fixed effects, echo estimates from prior models. Figure 4.5 displays predicted probabilities from this model of rejection as a function of co-authorship distance, with other variables held at their means.

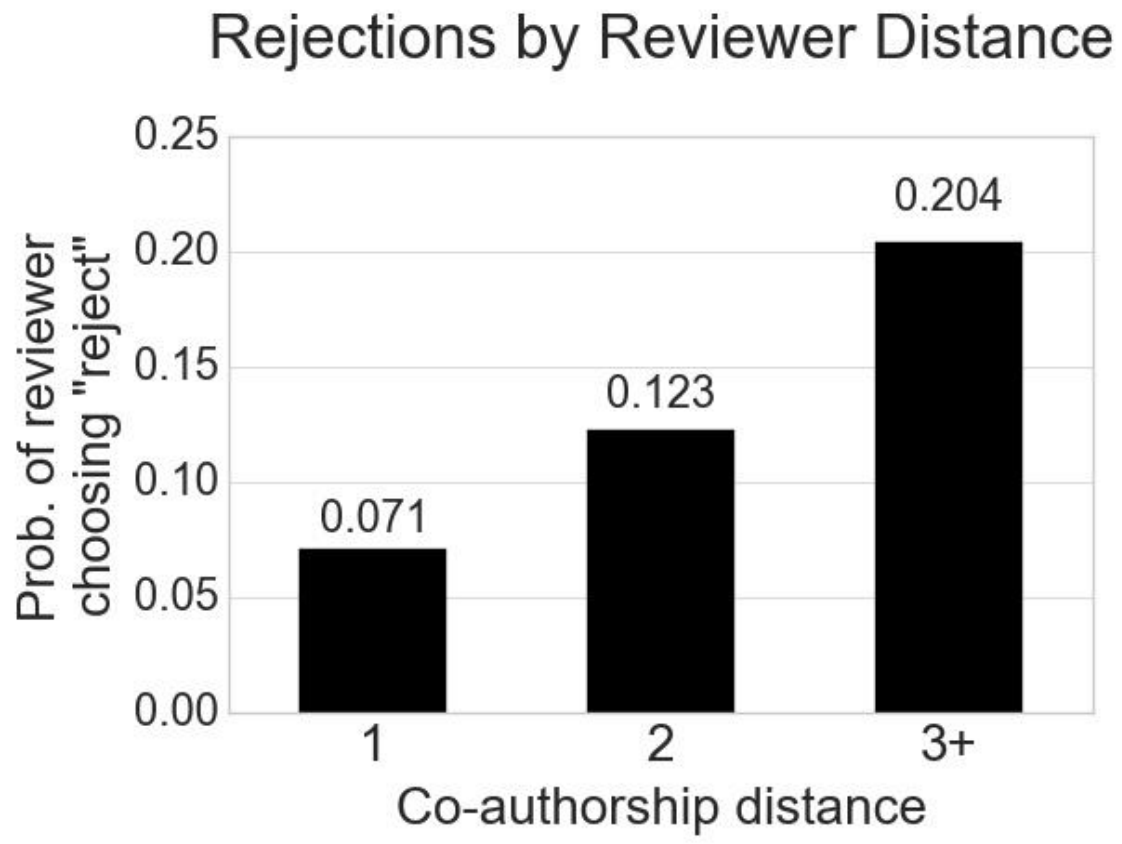

**Figure 4.5.** Predicted probabilities of reviewers recommending "Reject" as a function of co-authorship distance between the reviewer and author(s).

The probability of rejection increases steadily with co-authorship distance: close reviewers rarely reject manuscripts (7.1%), while the most distant reject more than a fifth (20.4%).

## 4.5. Mechanisms driving bias

Previous sections established that *PLOS ONE* reviewers, tasked with evaluating scientific validity alone, disagreed as often as reviewers for conventional journals and displayed a bias towards closely connected authors. What drives co-authorship bias? Insofar as it is followed, *PLOS ONE*'s validity-oriented review should rule out the mechanism of subjective review criteria. Indeed, the review process was designed specifically with this goal in mind. Nevertheless, we cannot rule out the possibility that reviewers fail to comply with instructions. In principle, editors should be able to identify non-compliant reviewers and either train them or exclude them from future invitations.

We now consider the two remaining mechanisms, nepotism and schools of thought. Section 2.2 argued that nepotism should be most salient in competitive, zero-sum environments. *PLOS ONE* is relatively non-competitive by design, however: it accepts 70% of submissions and any submission meeting its standards of scientific validity can be published. Another argument comes from equivocal outcomes in experiments designed to reduce bias directly by anonymizing scientific work before review. Such “blinding” experiments should make nepotism unlikely, as reviewers should be unable (or at least unlikely) to identify the author(s), and consciously or non-consciously deduce self-benefit. These experiments have been inconsistent in finding even small effects (Fisher, Friedman, & Strauss, 1994; Jefferson et al., 2002; Lee et al., 2013), however, suggesting either that reviewers are capable of guessing authors identities or that nepotism does not play a major role in decisions.

[18] Detailed estimates are available upon request.

An additional argument is based on distinguishing close, distant, and very distant reviewers. As discussed in section 2.2, close reviewers should have the highest incentive to review nepotistically, *i.e.* favor close connections on non-scientific grounds, and distant and very distant reviewers should have little to no such incentive. Our co-authorship measure enables us to separate reviewers into three groups: “close” (co-authors), “distant” (co-authors of co-authors), and “very distant” (co-authors of co-authors of co-authors or farther). We estimate the intensity of bias for these three reviewer types by estimating a regression identical to Model 4, except with co-authorship distances treated as a 3-level factor. Table 4.2 displays estimates from this model.

**Table 4.2.** Regression estimating the bias shown by close, distant, and very distant reviewers.

| | Dependent variable: *Review score* (1.0=Reject, 4.0=Accept) |
|---|---|
| *Co-author distance = 2* | $-0.144^{**}$ |
| | (0.058) |
| *Co-author distance = 3+* | $-0.232^{***}$ |
| | (0.058) |
| Controls: | |
| *Reviewer* h-*index* | -0.001 |
| | (0.002) |
| *Reviewer* n. *of co-authorships* | $-0.001^{*}$ |
| | (0.0004) |
| Manuscript FE | Y |
| Observations | 9,040 |
| $R^2$ (within-manuscript) | 0.008 |
| F Statistic | $5.637^{***}$ (df = 4; 2963) |

*Note:* Estimates from OLS regression. The dependent variable is Review score, which takes on values 1.0=Reject, 4.0=Accept. *, **, and *** indicate significance at the $p < .1$, $p < .05$, and $p < .01$ levels, respectively, with 2-tailed *t*-test. The $R^2$ is the within-manuscript variance explained.

As expected, distant (distance 2) and very distant (distance 3+) reviewers give substantially lower scores than close reviewers (distance 1, reference category). Most noteworthy, however, is the degree of favoritism exercised by distant versus very distant reviewers. Distant reviewers are $0.232 - 0.144 = +0.88$ more favorable than those very distant. Both reviewer types should have little to gain from nepotism, an assumption institutionalized around the world in policies that recuse only the most closely connected reviewers from reviewing. Yet even among distant reviewers, more distant reviewers are substantially harsher. Indeed, the difference in favoritism at each step in the co-authorship network is statistically indistinguishable - the difference between distant versus very distant reviewers is on the same order as that of closest connected versus distant, "arm's length" reviewers[19]. This pattern of bias intensity does not square with nepotism mechanism but is consistent with schools of thought: if differences in views on contested substantive matters vary smoothly with network proximity, distant reviewers should be more favorable than very distant ones.

[19] The calculation is available upon request.

These lines of argument are only suggestive. Non-compliance with review instructions and persistent (and possibly non-conscious) nepotism surely play a role in evaluation; there are too many reported experiences to claim otherwise. Nevertheless, reports also firmly support the arguments above in converging on the consistency of our findings with a schools of thought mechanism, the view that on contested scientific terrain, even well-intentioned reviewers will favor individuals who share the scientific view they themselves espouse and hold. We discuss policy implications from this finding below.

# 5. Discussion

A persistent critique of scientific peer review is that review decisions appear subjective. Instead of converging on the underlying quality of a scientific work, reviewers disagree with one another frequently and systematically favor work from those within their professional networks. We drew on literatures across the social sciences to identify and evaluate dominant explanations for how such patterns of disagreement and bias occur in science. Quantitatively oriented scholars, such as information scientists and economists, typically focus on nepotism and subjective review criteria. According to this view, a peer review system that evaluates work on criterion of scientific validity and minimizes competition between authors and reviewers should reduce disagreement and reduce or eliminate favoritism. On the other hand, historians, scholars of Science and Technology Studies, and other qualitatively oriented researchers highlight that at the research frontier, scientific validity itself is contested by schools of thought with distinct epistemic, especially methodological, commitments. According to this view, a validity-oriented peer review will do little to reduce disagreement or bias.

Our analysis of review files for neuroscience manuscripts submitted to *PLOS ONE* show that reviewers disagreed with each other just as often as reviewers in conventional review settings (inter-rater reliability = 0.19) that value more subjective criteria, including novelty and significance. Furthermore, we found the same pattern of favoritism: reviewers gave authors a bonus of 0.11 points (4-point scale) for every increased step of co-authorship network proximity. It bears emphasis that we observe these levels of disagreement and favoritism in a setting where reviewers are instructed to assess completed manuscripts on scientific correctness alone.

Such patterns of disagreement and favoritism are difficult to align with a view of the research frontier as unified by a common scientific method and divided only by distinctive tastes for novelty and significance. These patterns are instead entirely consistent with scholarship on “schools of thought.” According to this view, the research frontier is characterized by competing, epistemological cultures or schools of thought, members of which disagree on what methods produce legitimate claims and what assumptions are valid. The patterns we find are also consistent with nepotism - reviewers’ non-scientific considerations corrupt even straightforward assessments of validity. We presented several arguments that schools of thought play a major, if not primary role. First, incentives to review nepotistically are likely to be much lower in the modest-impact and non-zero sum setting of *PLOS ONE*. Second, policies designed to reduce bias directly, such as blinding reviewers from observing authors’ identities, have proven notably inconsistent in identifying even small effects (Jefferson et al., 2002; Lee et al., 2013). Lastly, as discussed in section 4.5, even distant reviewers who we expect to be non-nepotistic, give more favorable scores than very distant reviewers, who we expect to be equally non-nepotistic. Although not firmly conclusive, we believe this study provides strongest quantitative evidence for schools of thought.

## 5.1. Limitations

Our study has several limitations. First, the data are limited to a single discipline – neuroscience. Neuroscience combines methods and concepts from a variety of disciplines, so schools of thought may be particularly salient. How well our findings generalize to disciplines with more monolithic, shared epistemic cultures, such as economics, is unclear. Second, the uniqueness of *PLOS ONE* is both strength and weakness. Its review system and non-zero-sum acceptance policy make it particularly useful for disentangling mechanisms of reviewer decision-making. These same mechanism, however, may be difficult to generalize to more conventional and competitive review settings. Although the intensity of disagreement and bias in *PLOS ONE* suggest that it is not entirely unusual, there are few studies with which our estimates can be directly compared. It is thus possible that dynamics in *PLOS ONE* represent the "lower bound" for pathologies of peer review. Third, our analysis did not conceptualize or attempt to measure the true value or quality of manuscripts. Consequently, we cannot directly distinguish whether closely connected reviewers overestimate the quality of manuscripts or distantly connected underestimate it. We can only conclude that if the scientific validity of a manuscript can be conceptualized to have a single "true" value, the reviews of closely or distantly connected reviewers over- and/or under-estimate this value, respectively.

Fourth, the models we estimate here explain only a small proportion of within-manuscript score variation. This explanatory power is consistent with prior studies (e.g. Bagues et al., 2016) and may result from how editors select reviewers. Editors typically seek reviewers with substantial expertise but without close ties. The observed variation of expertise and connectivity are thus highly constrained. It is possible that in settings with larger variation in these quantities, biases would be more pronounced. Low explained variance may be interpreted in other ways as well. It is possible that in most cases, reviewers assess "normal science," with uncontested validity. In such cases, the review system may approach its idealized form – reviewers agree and hold few systemic biases. Conversely, contests of scientific validity may draw on epistemic views that are idiosyncratic, rather than organized within schools of thought that can be measured with formal relationships. In sum, disagreements in the evaluation of scientific validity are substantial, poorly understood, and continue to be an important area for research.

## 5.2. Policy implications

At the heart of this paper is an evaluation of a publishing experiment. *PLOS ONE* sought to improve the objectivity of peer review by designing a review system that minimized two intuitive causes of subjective evaluation: ill-defined and non-technical review criteria and nepotism. Accordingly, *PLOS ONE* evaluates manuscripts only on whether claims are valid and they minimize strategic considerations by publishing all work that meets this criterion. Our results indicate that, at least for multi-paradigm fields like neuroscience, this policy does not materially reduce disagreement in assessments or eliminate bias. The “tenacity” of these patterns suggests that schools of thought are a fundamental reality along the research frontier and should be taken into account by research organizations like journals and funding agencies.

The choice of a reviewer is typically conceptualized as a trade-off between expertise and bias (Bagues et al., 2016; Laband & Piette, 1994b; Li, 2017). One can choose (1) an expert reviewer who is biased toward her own research area, but can discriminate more effectively between high and low-quality work in this area or (2) a less expert reviewer who is unbiased but less effective or discriminating. In prospective evaluations of uncertain, future performance with reviewers possessing diverse expertise, this may well be the choice. Our work highlights that in assessments of concrete, already completed work and at high levels of expertise, reviewers do not uniformly favor or disfavor work within their research area. Instead, one faces the choice between (1) a “positive” expert reviewer from within the author(s)’ schools of thought or (2) a “negative” expert reviewer outside it (Travis and Collins 1991). Complicating the choice further, in many cases the difference between these experts will not be in how much they are corrupted by nepotism but where they fall on fundamental disagreements over whether the science is correct.

In practice, scientific organizations counter the possibility of bias with an intuitive policy: recuse reviewers from reviewing individuals to whom they are closely connected. The policy is a sensible response to the assumption that only close reviewers are corrupted by nepotism. Our work shows that even if nepotism is the mechanism driving bias, it extends beyond close connections: reviewers *without* close connections are biased relative to still more distant reviewers, and the intensity of bias is statistically indistinguishable as one reaches farther and farther in the co-authorship network.

The schools of thought mechanism points to another policy response to reduce bias. If reviewers from the same co-authorship cluster tend to share scientific views and favor each other’s work, fair reviews will require recruiting reviewers from diverse co-authorship clusters. Reviewers representing different schools of thought should be more able to recognize and make explicit to the editor or administrator the views and assumptions *not* shared, and their recommendations should reveal how crucial such differences are. Put in other words, editors and funders who value validity and diversity of published or funded output should also value diverse evaluators. In this way, scientific peer review converges on a pattern observed in many other evaluative contexts: from the selection of R&D projects (Criscuolo, Dahlander, Grohsjean, & Salter, 2016) to the hiring of candidates in elite service firms (Rivera, 2012) to policy choices by senators (Stanfield, 2008), diversity in outcomes requires diverse evaluators (Page, 2008, 2010).

Lastly, our study has implications for the interpretation of cumulative advantage so commonly observed in science and other social spheres (DiPrete & Eirich, 2006; Robert K. Merton, 1968). Cumulative advantage is usually thought to arise in settings where people lack complete information about the quality of potential choices and infer quality from existing status (Correll et al., 2017; Sauder, Lynn, & Podolny, 2012). Our findings highlight a mechanism that can give rise to cumulative advantage even in the absence of status signals. High status authors tend to have more connections: they train and co-author with more people and thus induce larger schools of thought (Frickel & Gross, 2005). An editor or funder who chooses reviewers based on expertise alone is thus likely to "sample" a reviewer from large schools of thought, and, unintentionally, solicit more favorable reviews. Conversely, lower status authors who belong to smaller schools would be assigned to reviewers from outside their schools. These unfavorable reviewers will lead to lower chances of publication, less status for the author, resulting in a downward status spiral.

In summary, we find evidence that distinct schools of thought appear to account for some, if not most, of the variance in reviewer outcomes previously attributed to subjective review criteria and nepotism. Our analysis is suggestive of the potential for research to more clearly model and measure scientific consensus and policies. To the extent that the research frontier is characterized by distinctive schools of thought with divergent epistemological commitments, our work shifts emphasis away from a mission to eliminate subjectivity through eliminating "corrupted" reviewers towards one that diversifies and balances it.

# Supplementary Materials

## A. Co-authorship network

We constructed the co-authorship network using the *Scopus* database. The name and institution of each author, reviewer, and editor was matched to his or her *Scopus* ID number and this number was queried to identify the individual's life-time co-authors. The query used *Scopus*' author search API (http://api.elsevier.com/content/search/author) and the "?co-author=" field. This query returned up to 179 co-authors for any particular individual. This artificial limit is unlikely to substantively affect our analyses because only 0.72% of scientists in the network have 179 or more co-authors. Figure A1 illustrates that the number of individuals with 179 authors is anomalously but not substantively high.

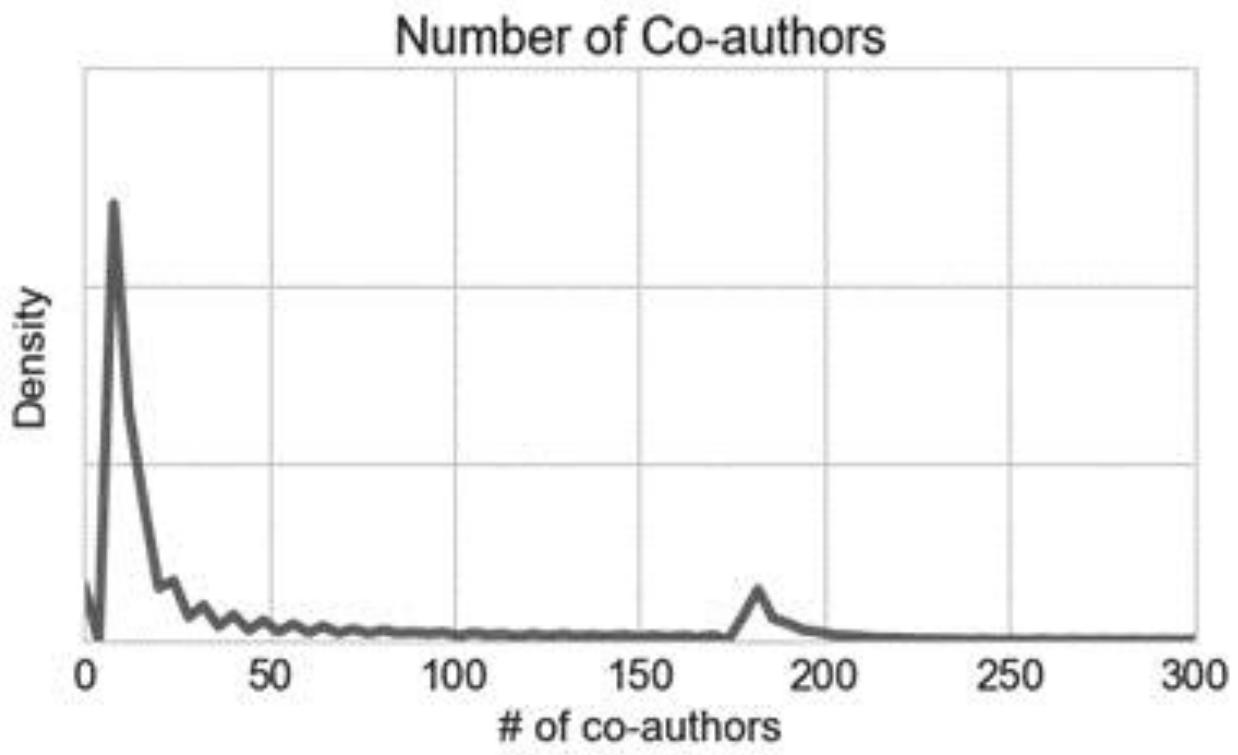

**Figure A1.** Distribution of the number of co-authors. The peak at 179 is a consequence of this being the maximum number of co-authors the *Scopus* API could return.

The network is constructed "one-level-out", with no measurement of co-authors of co-authors, using *PLOS ONE* authors, reviewers, and editors as seed nodes. The final network consists of 1,822,998 nodes and 4,188,523 co-authorship edges. Although the network is constructed only one-level-out, it is well connected. Each author, has, on average, 4.60 co-authors and 99.59% of the individuals are connected to one another by at least one path.

## Measurement error

Although the network includes nearly 2 million individuals and more than 4 million edges, it still leaves out an unknown number of relevant scientists and co-authorship ties. How does missing data affect the validity of measured co-authorship distances?

**Table A1.** Effect of missing network data on measured co-author distances.

| **Minimum co-authorship distance between reviewer and author(s)** | **Description** | **Expected effect of missing nodes and ties** |
|---|---|---|
| 1 | Reviewer has co-authored with at least one author | None |
| 2 | Reviewer shares co-author with at least one author | None |
| 3 | Reviewer shares co-author of co-author with at least one author | None |
| 4+ | Reviewer and author are separated in co-authorship network by at least 3 individuals | These measurements will, on average, be biased upward due to missing network nodes and edges. Two individuals who according to the measured network are relatively far from one another, e.g. co-authorship distance 5, may actually be at distance 3, if missing authors and co-authors were included in the network. 47.6% of the relationships in *PLOS ONE* are of length > 3. |

Distances 4+ are thus biased upward to an unknown degree. Consequently, we recoded these distances as 3. We use "3+" to refer to this recoded set together with the set of distances actually measured as 3.

Another major limitation of these co-authorship data is the lack of information on when these co-authorships were created, i.e. when the co-authored publication was published. Although the *PLOS ONE* data contain manuscripts submitted in 2011 and 2012, the co-authorship network takes into account publications published after 2012. An analysis of a random sample of 100 co-authorship links indicates that about 23% of these did not exist in 2012 or earlier. This measurement error acts to (erroneously) decrease the measured co-authorship distances between authors and reviewers relative to the distances that would have existed[20] in 2012.

One consideration limits the severity of such an error. Even a lively collaboration may not yield a publication for months if not years. Thus, it is likely that a co-authored publication published a year after 2012 reflects work that occurred months or years prior, particularly because *Scopus* does not index the pre-print archives relevant to neuroscientists. Consequently, many of the relationships formally recorded after 2012 were likely present many years earlier.

[20] For observed co-authorship distances of 1 (the two individuals are co-authors), in 23% of the cases the individuals were actually more distant in 2012. For observed co-authorship distances of 2, in (1.00 - 0.77 x 0.77) x 100% = 41% of the cases the individuals were actually more distance in 2012.

## B. Correlation matrix of variables used in the analysis

**Table 7.1.** Correlation matrix of variables used in the analysis.

| | **Review score** | **Co-author distance** | **Suggested** | **Reviewer's *n* of ties** | **Reviewer's *h*-index** | **Author(s)' *n* of ties** | **Author(s)' mean *h*-index** |
|---|---|---|---|---|---|---|---|
| **Review score (1.0=Reject, 4.0=Accept)** | 1.000 | -0.102 | 0.136 | -0.012 | -0.014 | -0.009 | 0.011 |
| **Co-author distance (minimum)** | - | 1.000 | -0.274 | -0.280 | -0.257 | -0.124 | -0.103 |
| **Suggested (=1 if author nominated)** | - | - | 1.000 | 0.087 | 0.135 | -0.022 | -.008 |
| **Reviewer's *n* of ties** | - | - | - | 1.000 | 0.752 | 0.099 | 0.051 |
| **Reviewer's *h*-index** | - | - | - | - | 1.000 | 0.050 | 0.031 |
| **Author(s)' *n* of ties** | - | - | - | - | - | 1.000 | 0.769 |
| **Author(s)' mean *h*-index** | - | - | - | - | - | - | 1.000 |